# Digging Deeper: Methodologies for High-Content Phenotyping and Knowledge-Abstraction in *C. elegans*

Nan Xu[1,*], Dhaval S. Patel[1,*] and Hang Lu[1]

[1]School of Chemical & Biomolecular Engineering, Georgia Institute of Technology, Atlanta, Georgia 30332, USA

[*]These authors contributed equally.

**Abstract**

Deep phenotyping is an emerging conceptual paradigm and experimental approach that seeks to measure many aspects of phenotypes and link them to understand the underlying biology. Successful deep phenotyping has mostly been applied in cultured cells, less so in multicellular organisms. Recently, however, it has been recognized that such an approach could lead to better understanding of how genetics, the environment, and stochasticity affect development, physiology, and behavior of an organism. Over the last 50 years, the nematode *Caenorhabditis elegans* has become an invaluable model system for understanding the role of the genes underlying a phenotypic trait. Recent technological innovation has taken advantage of the worm's physical attributes to increase the throughput and informational content of experiments. Coupling these technical advancements with computational/analytical tools has enabled a boom in deep phenotyping studies of *C. elegans*. In this review, we highlight how these new technologies and tools are digging into the biological origins of complex multidimensional phenotypes seen in the worm.

## 1. Introduction

One of the great drivers of biological research in the 20th century was the desire to understand how the information encoded in an organism's genome gives rise to its physical and behavioral phenotypes, collectively referred to as the organism's phenome[1]. However, the phenotype an organism displays is not merely a reflection of its genes but the integrated product of its genotype coupled with the environmental conditions and stochastic effects that the individual experiences before observation[2]. This interplay means that phenome (collection of phenotypes) of an organism contains a vast multi-dimensional set of observable characteristics, and thus in a population of individuals, the resultant phenospace is effectively infinite. Understanding the specific contribution of an individual's genome, its environment, and stochastic processes to its position in phenospace presents a formidable technical challenge. Fortunately, the pace of technological innovation in modern biology is beginning to provide the tools that are necessary for both quantifying these multi-dimensional characteristics and their underlying causes, opening up a new experimental paradigm known as deep phenotyping.

The laboratory animal *Caenorhabditis elegans* has many attributes that make it an ideal model system for deep phenotyping studies. For example, the worm is a small poikilotherm with an adult body length of ~1mm that feeds on bacteria. It has a rapid life cycle, going from egg to reproductive adult in ~3 days at 20°C with a deterministic developmental lineage[3]. These features coupled with the fact that the worm primarily reproduces as a self-fertilizing hermaphrodite mean that it is possible to grow large near-isogenic populations in highly controlled environmental conditions. Most importantly, at least ~38% of the protein-coding genes in the *C. elegans* genome have a human ortholog[4], which means that insights gained from deep phenotyping studies in the worm can inform us about human biology.

In this review, we will outline the technological and analytical developments that have enabled recent deep phenotyping studies in *C. elegans* (Section 2). We will then examine how these tools have yielded greater insight into the biology of complex phenotypes using several different examples (Section 3). Finally, we will offer a prospective outlook on the future of deep phenotyping experiments in *C. elegans* (Section 4).

## 2. Recent Development of Tools and Techniques that Enable Deep Phenotyping

Historically, phenotypic analysis of *C. elegans* was based on the manual measurement of easily scorable morphometric or behavioral features, such as alterations in body shape or defects in movement[5]. However, modern biological techniques are dramatically expanding the definition of phenotype[1]. For example, access to RNAseq and mass spectrometry is now more widely available and less cost prohibitive, enabling transcriptomic and proteomic descriptions of individuals. Similarly, advancements in both hardware and computational analyses have led to an increase in both the throughput of experiments and their informational content. In the remainder of this section, we examine three specific areas that have been particularly influential in the evolution of deep phenotyping studies of the worm.

### 2.1. Manipulating the *C. elegans* genome

*C. elegans* researchers have developed a sophisticated set of genetic tools with which to manipulate the worm's genome and illicit phenotypes, these include both forward and reverse genetic approaches[3]. RNA interference-induced knockdown (RNAi) of gene expression can be achieved by feeding worms bacteria expressing double-stranded RNA corresponding to the gene of interest[6,7]. This relative ease of RNAi has enabled multiple reverse genetic screens to uncover the phenotypes associated with various gene inactivations[8,9]. More recently, several groups have introduced CRISPR/Cas9 methods that are optimized for manipulating the *C. elegans* genome[10,11]. These methods allow for the rapid and efficient knock out of genes or introduction of fluorescent markers of either gene or protein activity.

### 2.2. Hardware employed in deep phenotyping

The hardware development that has enabled deep phenotyping studies in *C. elegans* broadly falls into two categories. The first is an improvement in the technology that allows manipulation of the worm and the second is an improvement in the technology that records the output of the experiment. Both categories have seen significant reductions in component costs and accumulation of many designs and proof-of-principle experiments, which increases the accessibility of the hardware needed for deep phenotyping experiments. We highlight the technical aspects in this section and give specific examples of using the hardware in section 3.

Manual handling of *C. elegans* can be very labor intensive and may act as a barrier to the scale and scope of an experiment. However, the small size of the worm makes it very easy to manipulate using microfluidics[12]. In the last decade or so, the microfluidic devices used by the *C. elegans* research community have been fabricated from polydimethylsiloxane (PDMS). The fabrication of devices with is this material is relatively cheap and straightforward, and PDMS is non-toxic to worms. It is also optically transparent, which makes it compatible with many forms of microscopy used to study the worm. Its elastic properties allow for simple on-chip valves that can control the flow of fluid containing the worms, which enables automation of worm handling leading to increases in sample throughput[12-14]. In addition, microfluidics can also be used to tightly control the microenvironment surrounding the worm within the device, something that is much harder to do on an agar plate. Typical device designs that are widely used for imaging-based or behavior-based experiments include multi-chamber arrays[13,15] (Fig 1A), sorting devices[16-18] (Fig 1B), and arena devices[19] (Fig 1C), specific examples of the use of these devices are described in section 3.

The majority of deep phenotyping experiments in *C. elegans* rely on some form of optical readout of the data being recorded, often involving automated image capture. Studies that focus on high-spatial resolution examination of specific cells, tissues, or regions of individual worms are now more accessible to researchers due to the increased availability of epi-fluorescent and confocal microscope systems[20-23]. Similarly, many high-content behavioral studies use imaging systems that allow continuous capture of worms at relatively low magnification for the longitudinal tracking and quantification of multiple whole-animal level phenotypes[24]. These systems typically utilize dark-field or transmission imaging to enhance the contrast of the transparent worms. The cost of components needed for dark-field illumination setups has fallen precipitously in recent years, which has led to an explosion in automated systems designed to track freely moving worms[24,25].

### 2.3. Computational Techniques for Deciphering High-Content Phenotypic Information

A consequence of the hardware-related gains in throughput and information recording is a dramatic increase in the data produced by deep phenotyping assays compared to more traditional experimental pipelines. These experiments often produce volumes of data that are well beyond the analytical capability of an individual human being, which in turn has spurred the development of efficient computational tools to parse it. Given the ubiquity of microscopy in *C. elegans* research, the primary set of computational tools that have enabled deep phenotyping studies are various forms of image processing algorithms. A large body of literature exists on methods for image segmentation and reconstruction to extract the morphological features not only in *C. elegans* but also in other species such as Drosophila and mouse[26], on labeled synapses[16,27-30], cell nuclei[31-35], neuronal structure[36-40], and cells in general[41]. It should be noted that not all of these image processing techniques have been applied in deep phenotyping yet; we envision that many of these tools, when adapted for specific applications, could enable high-content experiments and thus deep phenotyping in the future. This will likely require the image processing and biological research communities to establish close collaborations, as there are still significant barriers in translating knowledge between fields and mapping the technical know-how to specialized application details.

Beyond image processing, additional computational approaches, include methods in stochastic processes[42-44], information theory[44-46], and machine learning[16,17], have been introduced into the field in recent years. One important class of techniques is the dimensional reduction of features to probe the structure of the data[17,47-53]. Several unsupervised learning approaches have been applied for this purpose, such as principal component analysis (PCA) and clustering, whose use was inspired by their application in earlier non-worm studies in gene cluster identification[54], cell identification[55], and human brain activity[56]. More recent efforts that focus on feature classification employ machine learning methods such as support vector machine (SVM) or logistic regression models[15-17]. A third class of analytical tools that focus on deciphering the transition processes of biological states employ stochastic processes and information theoretical methods. An example of the use of these tools is the study of *C. elegans* locomotion[44,45,57], which involves continuous transitions in worm posture. The use of these tools is driven mainly by the need to process complex information that usually escapes human visual detection and imagination, and to avoid human bias. As we see more data gathered using advanced experimental tools and the complexity of the data increases over time, these computational and theoretical tools are likely to gain more prominence in the biological discovery process.

## 3. Recent Applications of Deep Phenotyping

Given that the hardware and computational tools to perform deep phenotyping experiments in *C. elegans* now exist, we next turn to question of how these technologies have been used and what they have taught us. In this section, we review a range of recent studies covering diverse areas of worm biology.

### 3.1. Embryogenesis/lineage tracing

*C. elegans* is one of the few metazoans for which the entire somatic cell lineage can be traced from single-cell embryo to adult. However, tracing the lineage of cells in the developing worm embryo is exceptionally challenging due to the rapidity of cell division and morphological similarity of the cells[58]. The relative complexity of identifying phenotypic aberrations in the embryonic lineage compared to that of the post-embryonic cells has meant that the mechanisms governing embryonic cell division and differentiation have been harder to elucidate. The advent of 4-Dimensional imaging systems[59,60] removed the need for manual observation of embryogenesis but not curation of the resultant images into lineages. Deep phenotyping is still possible using manually curated data, for example, a study involving a whole genome RNAi screen identified 661 genes involved in the early embryogenesis[61]. A subsequent study integrated transcriptional, protein-interaction and visual phenotypic data of these 661 genes to create predictive models of cellular events during embryogenesis[62] (Fig 2A). The use of transgenic worms that ubiquitously express fluorescently-tagged histone proteins has enabled the development of automated tracing algorithms that track the embryonic lineage through to the 350-cell stage[63-68] (Fig 2B).

With automated image collection and curation in place, it is now possible to examine different aspects of embryonic cell division. Several studies used automated lineage tracing to identify the precise cellular expression patterns of known embryonic genes[69-71]. Automated cell lineage tracing has allowed for the construction of a single cell resolution atlas of gene expression revealing when and where transcription factors are expressed in the developing embryo[70,71]. It has now become possible to ask how different genetic perturbations affect embryogenesis. Early examples of this include identifying the subtle roles of a single transcription factor[72] and the distinct roles of highly-related/recently duplicated genes[73] in defining different aspects of the embryonic lineage. More recent advances have demonstrated the roles of several hundred genes in cell fate choice[74,75], the regulation of asynchronous cell division[76]. A more limited screen of chromatin regulators has also revealed distinct roles for several chromatin modifying complexes during embryogenesis[77]. Together these deep phenotyping studies are revealing the genetic programs and molecular mechanisms that specify how a single fertilized oocyte becomes an embryo containing a complex collection of differentiated cell types.

In contrast to tools available for deep phenotyping of the embryonic development, the ability for high-throughput examination of the cell lineage of post-embryonic worms has been lagging. However, recently Keil *et al*. have demonstrated a microfluidic device that allows the long-term culture of *C. elegans* larvae coupled with the ability to routinely immobilize the animals in order to take high-resolution images with a variety of microscopy techniques[78]. Using this platform, the authors demonstrated the ability to image the animals from the first larval stage (L1) through to egg-laying adult, while examining vulva precursor cell development, apoptosis during larval molting, neuron differentiation and neurite outgrowth. This type of platform coupled with future developments in post-embryonic phenotyping is likely to lead to a complete description of the biological events involved in the development of a multicellular animal.

### 3.2. Automated high throughput genetic screens

One of the most potent aspects of *C. elegans* as a model system has been the ability to carry out forward genetic screens in order to identify mutations that affect all possible elements of a gene. However, most genetic screens rely on visually identifiable phenotypic differences from the control, which inherently limits the ability to identify mutations that have weak or non-obvious phenotypes but still provide valuable information on a gene's function. The power of deep phenotyping in identifying subtle mutants that may be missed by visual inspection was demonstrated in automated screens to find mutations affecting synaptogenesis[16,17,79]. In particular, using a microfluidic sorting system[16] (Fig 1B), hundreds to thousands of worms can be continuously imaged, processed, and sorted in real time. An online image processing algorithm based on support vector machine (SVM), inspired by the recent success of using supervised learning in various computer visions applications, was developed to classify multidimensional features of synapses on the fly[16,17]. In addition, a stepwise logistic regression model was used to measure the degree of phenotypic deviation of mutagenized worms from wild-type[16,17]. Clustering of the multidimensional features derived from the quantification of the fluorescent synaptic marker in the worms from the screen reveals the phenospace of the entire mutational spectrum[16,17] (Fig 2C). These clusters indicate where each new mutant resides in phenospace, allowing the inference of their potential genetic relationships. This study offers a powerful example of how deep phenotyping integrates high-throughput hardware with computational tools to provide mechanistic insights that would not have been possible for a human observer.

### 3.3. Measuring 'whole-brain' neural activity

*C. elegans* is an excellent model system for examining how nervous systems encode environmental information and modulate animal behavior. However, measuring the activity of neurons as the worm responds to specific stimuli is a major technical challenge. The recent development of genetically encoded calcium indicators has allowed the imaging of neuronal activity, ranging from a single neuron all the way up to the entire 'brain' in the *C. elegans* nervous system (for a summary, see [80]).

'Whole-brain' imaging is an inherent form of deep phenotyping, as it involves long-term observation and analysis of a large number of neurons. The methodology has only recently become feasible in *C. elegans* because of advancement in speed and sensitivity of cameras and research and commercial success in new microscopy platforms. Several studies have successfullly characterized whole-brain dynamics under various experimental conditions. Kato et al.[81] used a microfluidic imaging platform[51] to carry out whole-brain recordings from immobilized worms, which reveals that the evolution of network dynamics among neurons are directional and cyclical. They showed that different phases in this cyclical activity regulate motor commands that drive certain locomotory behaviors. Using a modified commercial spinning disk confocal system, Venkatachalam et al.[82] developed a whole-brain imaging platform and studied representations of sensory input and motor output of individual neurons upon thermosensory stimulus in freely moving worms. Similarly, using a simultaneous worm tracking and whole-brain imaging system, Nguyen et al.[50,83] recorded whole-brain activities of freely moving without stimulation; in a related effort Nguyen et al.[83] showed a machine-learning approach to track neurons in the freely moving heads, which is an important step towards robustly and automatically analyzing such large sets of dynamical data . Most recently, Nichols et al.[84] investigated the global brain dynamics during lethargus, a sleep-like state in the worm. This study showed that global brain activity becomes quiescent during lethargus; interestingly, however, certain neurons remain active as these cells promote the establishment of the quiescent state. By examining the whole-brain dynamics, this work demonstrated that the transition to the 'wakeful' state is carried out by the re-establishment of activity in specific neurons that then drive global brain dynamics back to the aroused state[84].

From the above examples, it is clear that there is much to be gained by studying the entire nervous system, rather than analyzing a specific subset of neurons, as not all of the neurons nor their functional connections in the neural circuits governing various sensorimotor behaviors have been identified. These examples also point to many opportunities for future theoretical and technological development for analyzing and understanding such complex and dynamical systems. For instance, better imaging systems that allow coupling of other experimental techniques (e.g., microfluidics or optogenetics) and better/faster tracking algorithms and the automatic identification of the neurons are still needed. Further, better theories may be needed for interpreting the large volumes of curated data in the future to make sense of how the brain processes information and makes decisions.

### 3.4. Behavioral analysis

One of the original motivations that drove the development of *C. elegans* as a model system was the desire to understand how the nervous system of an animal gives rise to all the behaviors it elicits[5]. A recent extensive review by McDiarmid et al.[25] lays out the history and biological significance of behavioral studies of *C. elegans*. Recording behavioral phenotypes is technically non-trivial as worms exhibit rapid changes in direction and posture. Here we discuss briefly recent improvements in worm tracking hardware and software and show how deep phenotyping studies could reveal mechanisms underlying emergent behaviors.

The rapidity of changes of behavior in freely moving worms makes it nearly impossible for a human observer to record all events in real time; thus the majority of behavioral studies employ some form of automated image capture. The earliest studies were able to record the spatial position, speed, and turning rate of individual worms[42]. However, more recent tracking systems are capable of offering more comprehensive descriptions of worm behavior. Worms tracking systems fall into two categories: single-worm trackers for high-resolution analysis of individual behavior[19,85-87] and multi-worm systems for higher throughput population-level studies[88-92] (Fig 3). These systems have been used to elucidate the behavioral genetics of several sensory modalities. Explicit analyses of different behavioral features of thousands of animals from 239 genotypes revealed uncovered 87 genes, including those in sensory function and the Gαq signaling pathway, involved in worm locomotion and predicted 370 genetic interactions among them[93]. Similar studies have identified genes involved in thermotaxis[47], chemotaxis[88,94] and mechanosensation[88]. Most of these studies were performed on agar plates; in contrast, the development of microfluidic arenas allows for precise spatiotemporal control of the chemical environment revealing behaviors are not observable in plate-based experiments[95].

In addition to the improvement in worm tracker hardware, development of new algorithms has led to a more comprehensive description of worm behaviors. For example, PCA has been used to decompose the postural space of worms into eigenvectors referred to as “eigenworms”[53]; surprisingly, the postural space of locomoting worms on agar plates is low dimensional - superposition of four of eigenworms is sufficient to describe the majority of the worm’s locomotory postures. This analysis, which dramatically reduces the complexity of quantifying behavioral patterns, has been built into many worm tracking systems[52,53,86-88]. The approach has also enabled studies of behavioral dynamics[52,86]. Brown et al.[52] developed a dictionary of behavioral motifs for both wild-type and 307 mutant strains, which enables clustering of mutants into related groups, a task that would have been nearly impossible to do by eye. Similarly, Yemini et al.[86] established a comprehensive behavioral recording and curation system and developed an extensive phenotypic database of locomotory behaviors for a large number of strains, including multiple alleles of the same gene as well as some double and triple mutants. The growing dictionary[52] and database[86] have allowed users to uncover subtle behavioral phenotypes for mutants that could not be discerned by manual observation, underscoring the importance of deep phenotyping pipelines.

Understanding how these behaviors emerge from the integration of information about the external environment and the worm's own internal state remains limited. This problem has been studied extensively in *C. elegans* foraging behavior[44-46,96-98]. The behavioral state transitions a worm undergoes, as well as the informational value of the food it encounters while foraging can be formulated mathmatically[44,45]. For example, behavioral state transitions can be predicted by using the probabilistic transitions in a Markov process[44,57], and information about food can be evaluated by information theoretic measures such as mutual information[44,45]. These models can be used to predict the worm's response to food encountered while foraging[44,57]. We anticipate that when such predictive models are integrated into deep phenotyping studies, it will lead to a greater mechanistic insight into the genetics of foraging behavior.

Increasing throughput is essential to the deep phenotyping of foraging behavior. A recently developed microfluidic imaging system, WorMotel[99], is a multi-well PDMS platform that uses a chemo-aversive moat to trap animals to their specific well, which allows for highly parallelised monitoring of individual worms under uniformly controlled environmental conditions This platform has been used to examine the relationship between the roaming (active) or dwelling (sedentary) behaviors and food abundance[96,97]. These studies show reveal the biological complexity of foraging behaviors. For example, are serotonin produced by the ADF neurons promote roaming while serotonin produced by the NSM neurons promote dwelling[97]. Other biological amines such as dopamine promote dwelling[96], while octopamine is involved in roaming behaviors[97].

Similar parallelization has also been demonstrated on plate-based foraging assays through the use of multiple cheap cameras[46]. Stern et al. examined roaming and dwelling behavior in single worms continuously across their development[46]. The authors found that the pattern of behaviors varied across the different developmental stages and also between the onset and exit phases of each larval stage in a reproducible way. A suite of neuromodulators is responsible for the regulation of these behavioral patterns[46]. Interestingly, by tracking individual worms throughout their development, the authors also found that there was significant inter-individual variation in roaming behavior. Even though the worms were derived from isogenic populations, some animals consistently roamed less across all developmental stages while other consistently roamed more[46]. Quantifying inter-individual variation in any phenotype and understanding it biological origins is challenging, given the large numbers of individuals that need to be surveyed. Deep phenotyping offers an integrated way of achieving high-throughput experimentation with comprehensive behavioral analysis.

### 3.5. Measuring aging and age-related decline

Over the last 30 years, there has been a concerted effort among scientists to understand the biology of aging. *C. elegans* is the premier model organism for the study of longevity, due to its short lifespan and powerful genetics[100]. The majority of lifespan studies in *C. elegans* are performed manually by periodically examining animals maintained on agar plates under a stereomicroscope for either spontaneous or stimulated movement[101]. These manual experiments place constraints on the types of phenotypic and scale of demographic data that can be obtained by a human observer. Deep phenotyping technologies offer a more efficient and cost-effective way of collecting high-throughput and high-temporal resolution lifespan data.

The first push towards automating aging studies in *C. elegans* used commercial grade scanners to capture time-lapse data on worm movement on agar plates[102,103]. In these scanner-based systems, death is defined as a persistent absence of movement, as there is no ability to perform stimulation on the worms to test for induced movement. Stroustrup et al. used their automated system, termed the Lifespan Machine[103], to examine the demographic features of large populations[104]. The authors found that at the population-level aging appeared to scale temporally between groups that had very different mortality rates due to a variety of different factors. This scaling implied a single effective rate constant of aging in *C. elegans* and that interventions that altered longevity did so by changing this rate[104].

These scanner-based lifespan studies are unable to gather data at the level of the individual, as the temporal resolution of the scanner does not allow the tracking of each worm till it dies[103]. As a result, these systems cannot assess the level of inter-individual variation in lifespan within a population, which can be dramatic even for isogenic worms raised in identical conditions[105]. Two groups have recently demonstrated robust high-throughput automated acquisition of lifespan data from individual animals[106,107]. Pittman et al. use a polyethylene glycol hydrogel (PEG) substrate that is seeded with spots of *E. coli* as a food source for developing embryos that are individually placed in a single spot, once the embryos are deposited the substrate is sealed using a layer of PDMS[108]. The PDMS acts as a barrier confining each animal upon hatching to its local spot of *E. coli*. The PEG substrate can then be imaged using a wide array of microscopy platforms, which allows the tracking of multiple features of the development and lifespan of each worm individually. Zhang et al. used this system to show that when tracked at the level of the individual, aging does not display temporal scaling, with long-lived animals displaying an extended decrepitude phase the authors termed ‘twilight’ compared to short-lived individuals[107]. The WorMotel platform[106] (see Section 3.4.2), can record both spontaneous and stimulated movement, which is evoked by the use of a brief pulse of blue light, to assess whether a worm is alive or not. Using this system

Churgin et al. also demonstrate that short-lived and long-lived individual lifespans from an isogenic population do not temporally scale[106]. The differences seen between a population-level measure of longevity and those that account for inter-individual variability demonstrate the increasing power of deep phenotyping technologies.

Dietary restriction (DR) is one of the most robust interventions that has been shown to extend lifespan in an evolutionarily conserved manner[109]. There is considerable interest in finding drugs that can induce the phenotypic effects of DR without the concomitant need for drastic diet alteration in humans. Using the scanner-based lifespan machine[103], Lucanic et al. designed a high-throughput screen designed to identify potential DR mimetics in *C. elegans*[110]. The authors screened through a library of 30,000 small molecules, identifying 57 compounds that repeatedly extended the lifespan of treated animals versus control worms. Several of these compounds were structurally related, containing a nitrophenyl piperazine moiety, and further analysis of the most potent of these, NP1, suggested that it extended longevity by inducing DR-like effects[110].

One of the reasons for the recent rise in interest in the study of aging is that the incidence of both cognitive impairment and neurodegenerative disease increases with old age, which imposes a significant societal cost as the proportion of elderly individuals relative to those of working age in the general population continues to grow[111]. There is strong desire with the research community to develop treatments that can counteract the debilitating neurological effects of aging. *C. elegans* also displays age-related declines in cognitive ability and the morphological structure of its nervous system[112]. Bazopoulou et al. have developed a microfluidic platform to monitor the calcium responses of a specific polymodal neuron, ASH, as it ages[113]. They then used this system to conduct a pilot screen to identify small molecules that could delay the age-related decline of ASH activity. Several molecules from a panel of 107 FDA-approved compounds delayed the decline of calcium responses in ASH during aging, the most potent of these were Tiagabine and Honokiol[113]. This study demonstrates the power of deep phenotyping technologies to acquire dynamic readouts, such as neural activity, in a high-content manner.

### 3.6. Drug discovery and small molecule biology

Deep phenotyping is increasingly being embraced by researchers in the field of drug discovery as it allows rapid screening of the pleiotropic effects of different molecules. Given the ease of culture and amenability of *C. elegans* to high-throughput experimentation, the worm has been used to model many different human diseases[114,115]. While many researchers are developing methodologies to use these *C. elegans* disease models to speed up drug discovery, these platforms cannot yet be considered as examples of deep phenotyping as they tend to analyze single and relatively simple phenotypes. However, it is worth noting that as they serve as the as signposts for where we hope truly deep phenotyping platforms will be able to take drug discovery screens in model organisms.

A *C. elegans* model of cirrhosis, a mutant human α1-antitrypsin (ATZ) fused to GFP aggregates within the ER of worm intestinal cells replicating the phenotype seen in diseased hepatocytes[115]. In this study, a high-resolution plate reader was used to rapidly screen a commercially available compound library, yielding 33 compounds that decreased the rate of GFP-aggregation within the worms' intestinal cells[115]. More recently, a high-throughput genome-scale RNAi screen of ATZ model worms was performed a to find gene inactivations that alter the intestinal GFP-aggregation by the same group[116]. RNAi of 100 genes led to decreased levels of GFP-aggregation in the ATZ worms. An *in silico* approach then identified drugs that are known to target the mammalian orthologs of the worm genes and tested them for rescue of the GFP-aggregation phenotype. Poly-glutamine expansion (polyQ) diseases, such as Huntington's, are another example of a protein aggregation disorder that has also been successfully modeled in transgenic worms[117]. Recently, a new microfluidic system was unveiled for high-throughput drug discovery using the *C. elegans* polyQ model as a demonstration of the potential of the platform[118].

Using this platform, Mondal et al. rapidly screened ~100,000 worms through their device testing 983 FDA-approved compounds for their ability to reduce YFP aggregation in a high-polyQ strain[118]. The screen resulted in four compounds that had a statistically significant reduction in protein aggregate formation.

Application of deep phenotyping technologies may also enable the study of widespread but often neglected diseases in a cost-effective way. Parasitic nematodes are thought to infect a billion people worldwide and are also a significant source of infection in many animals and plants that humans are dependent on for food and their livelihoods. However, development of anthelmintic drugs has not kept pace with the acquisition of drug-resistance by these nematodes, and so there is an urgent need for new therapeutics (see references in [119,120]). Parasitic nematodes are difficult to work with directly due to the need to grow them in a host system, so *C. elegans* has become a model organism for anthelmintic toxicology. The WormScan platform, which uses a consumer-grade flatbed scanner as its basis, can simultaneously measure the mobility, brood size, body size and lifespan of worms either on agar plates or liquid culture[102,119]. This system was used to screen through 26,000 compounds resulting in the identification of 14 potential anthelmintic compounds. The INVAPP/Paragon assay uses a similar liquid culture strategy for screening for anthelmintic compounds[120]. This system also utilizes a 96-well plate format for culturing worms in liquid. However, the automated image capture relies on a high-frame-rate camera instead of a scanner[120], giving this system increased sensitivity in the detection of drug-induced motility defects. A proof-of-concept drug screen using this system with a 400-compound library identified 14 molecules that impaired *C. elegans* growth[120]. A separate compound library screen against the parasitic nematode *Trichuris muris* using INVAPP/Paragon uncovered an entirely new class of promising anthelmintics[121]. These deep phenotyping-based studies of *C. elegans* allow researchers to rapidly identify potential anthelminthic drugs and their mode of action without the need to perform the complex in-host assays required to study parasitic worms.

Small molecule screens are not only a tool to identify therapeutic chemicals, but they can also be used to study the biology of genetic pathways. The *C. elegans* gene *skn-1* is a transcription factor that regulates the worm's response to oxidative and xenobiotic stress[122]. Leung et al. performed a screen for small molecule activators of the SKN-1 protein by using a plate reader to measure the induction of a GFP-based transcriptional reporter of the *skn-1* target gene *gst-4*[123]. These authors then subsequently demonstrated the ability of their system to perform an ultra-high-throughput screen for inhibitors of SKN-1 amongst a compound library containing over 364,000 small molecules[124]. This screen resulted in 125 that specifically lowered the fluorescent signal via inhibition of the *gst-4::gfp* reporter, suggesting that these molecules were SKN-1 inhibitors[124]. A similar strategy has been used to study *C. elegans* male-specific linker cell undergoes cell death just after the molt between the fourth larval stage and adulthood[125]. The death of this cell is non-apoptotic as it is independent of caspases; however, its death shares many morphological features with other non-apoptotic cell death observed in vertebrate development[125]. Schwendeman and Shaham performed a proof-of-concept small molecule screen to identify potential inhibitors that may shed further light on the biology of this phenomenon[126]. A screen of 23,797 compounds using a laser scanning cytometer resulted in the identification of six compounds that caused persistence of the linker cell by inducing some form of global developmental delay in the worms that was rescuable upon removal of the worms from the drug[126]. Together these studies demonstrate how the use of deep phenotyping technologies could enable the quantitative measurement of multiple morphometric traits to gain new insight into biology.

## 4. Future outlook

In this review, we have highlighted many recent conceptual and methodological developments for deep phenotyping, specifically using *C. elegans* as a model system. From the success of these studies, it is clear that by measuring many aspects of the morphology, functional output, or behavior of cells, circuits, tissues, and individual animals, we can expand the scope of biological studies. Furthermore, by using appropriate mathematical and statistical tools, we can better understand their underlying biological mechanisms. We believe that these approaches will become more ubiquitous as improved microscopy and other experimental tools and analytical pipelines using advanced computational and theoretical techniques become more accessible. While many of these tools might be initially developed for *C. elegans*, their general utility will apply to a broad range of biological systems. We predict that with future integration of efforts in different disciplines (e.g., biology, engineering, and computational sciences), the ability to link phenotypes to genotypes, environmental conditions, and stochasticity will significantly accelerate.

**Acknowledgements**

We thank K. E. Bates, D. A. Porto and T. Rouse for their suggestions on relevant literature for this review.

## Figures

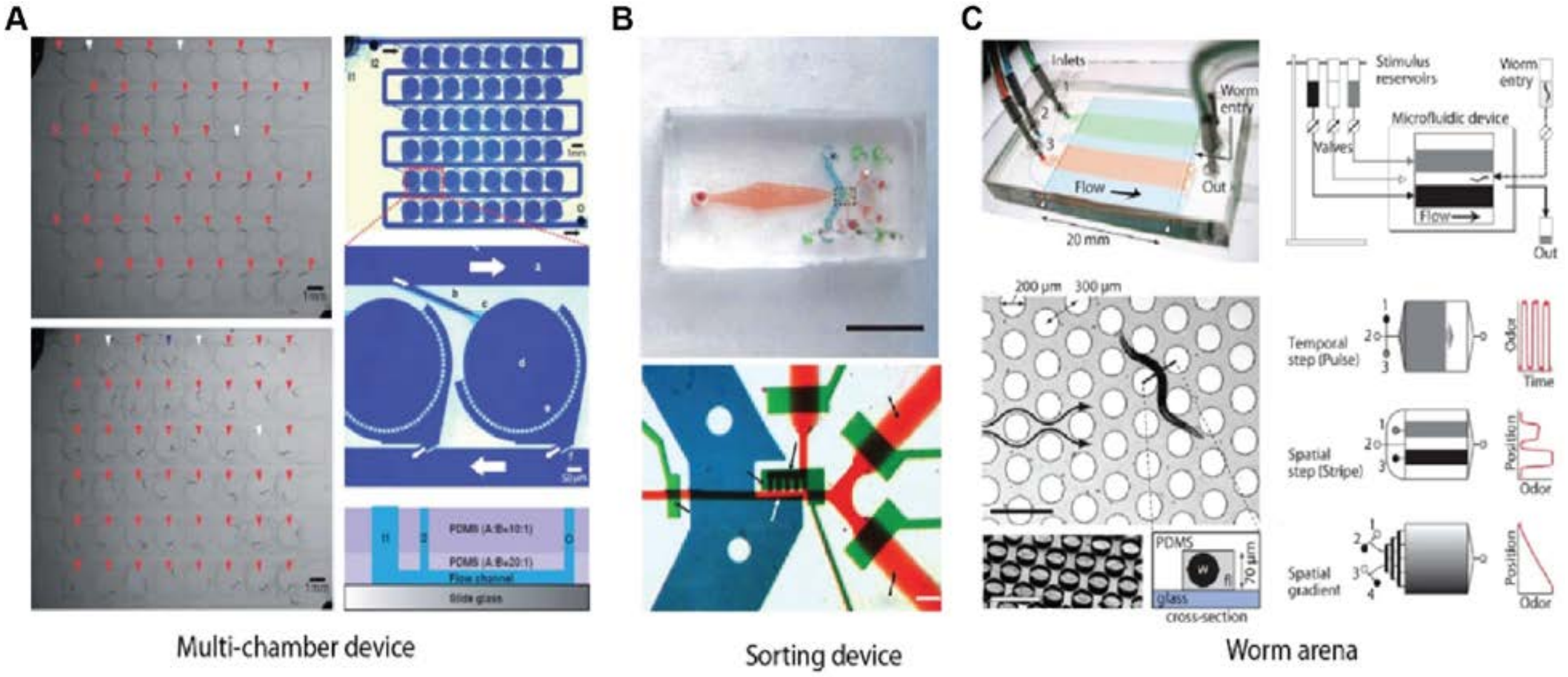

Figure 1. Microfluidics enables high-throughput experimentation in *C. elegans*. Examples of microfluidic devices routinely used in worm deep phenotyping studies. **A.** Multi-chamber arrays for simultaneously studying large numbers of individual worms (reproduced from [15]) **B.** Sorting devices used to rapidly isolate worms with specific characteristics (reproduced from [18]) **C.** Arena devices for behavioral assays (reproduced from [88])

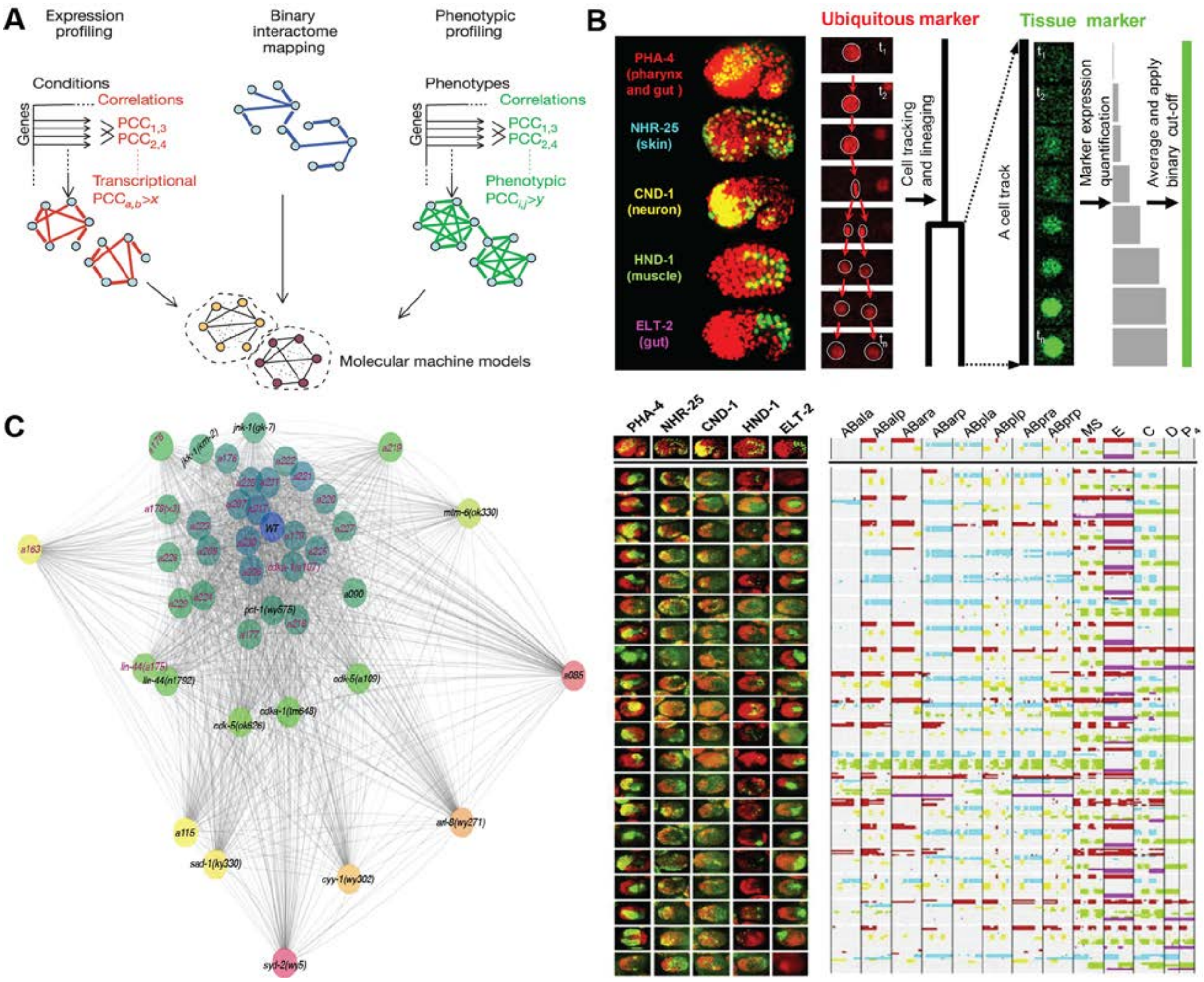

Figure 2. Deep phenotyping is a tool to produce informationally rich datasets. **A.** Integration of transcriptomic, proteomic and phenotypic data can be used to create models of cellular events in early embryogenesis. (reproduced from [62]). **B.** Use of two-color imaging can aid automated embryonic lineaging (upper panel) which can then be used to create spatiotemporal maps of gene expression (lower panel) (reproduced from [75]). **C.** Deep phenotyping can reveal a broader swathe of the phenotypic spectrum than traditional screens. Subtle mutants (labelled in pink) lie closest to wild-type (WT), while most previously identified mutants are the farthest from WT (reproduced from [17]).

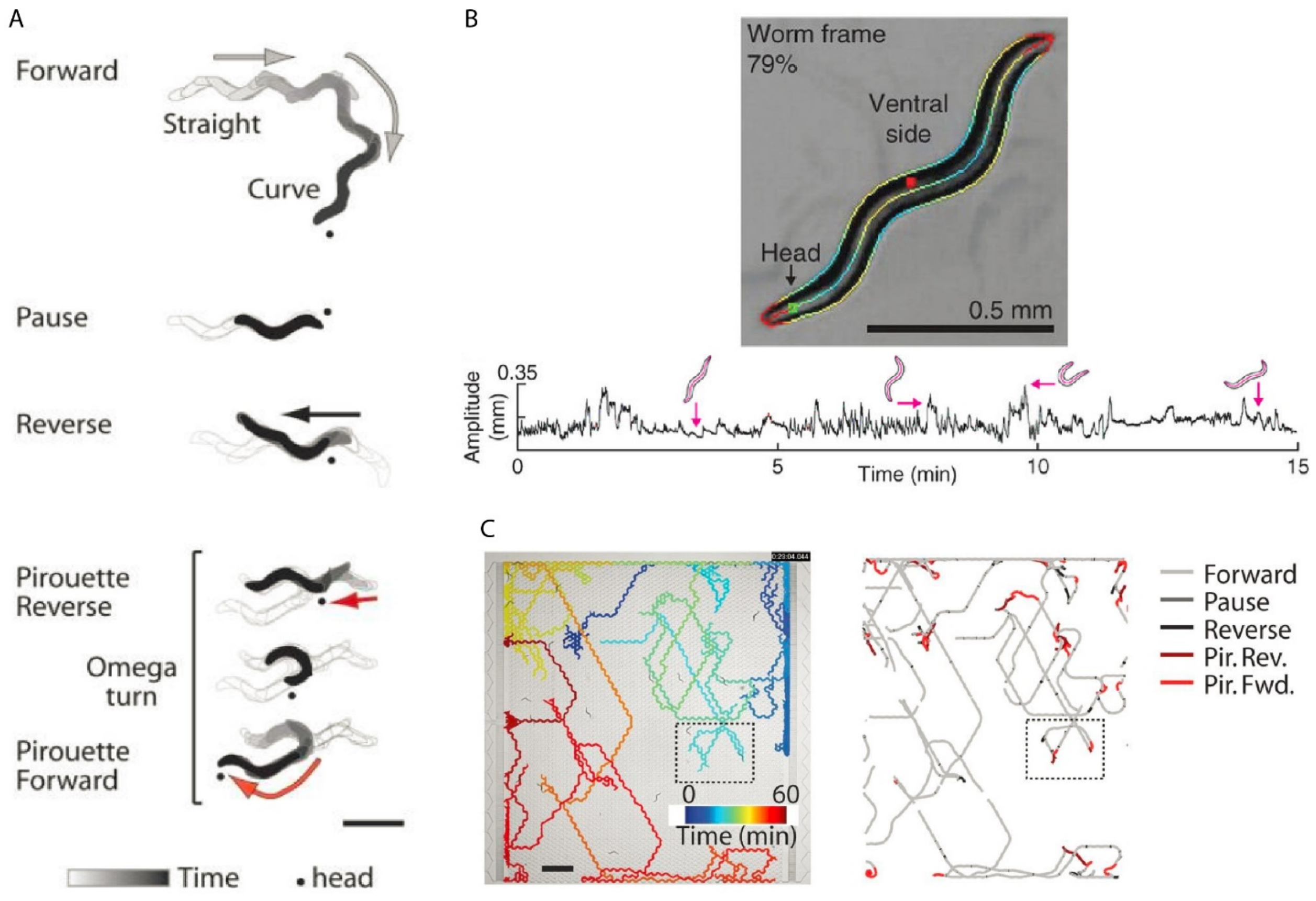

Figure 3. Behavioral deep phenotyping of worms using automated trackers. **A.** Behavioral repertoire of *C. elegans* that can be automatically tracked. **B.** Example of a single worm's behavior during an experiment. **C.** A multi-worm tracking system developed for tracking behaviors in a group of worms. (**A.** and **C.** are reproduced from [88], **B.** is reproduced from [93].)